\documentclass[aps,  showpacs]{revtex4}

\begin{document}
\title{Spintronics: Maxwell-Dirac theory, charge and spin}
\author{S. C. Tiwari \\
Department of Physics, Institute of Science, Banaras Hindu University,  and Institute of Natural Philosophy, \\
Varanasi 221005, India }
\begin{abstract}
The nature of spin current and the separation of charge current and spin current are two of the fundamental questions in spintronics. For this purpose the classical limit of the Maxwell-Dirac theory is investigated in the present contribution. Since the Dirac equation reduces to the Weyl equation for massless particles, a vortex solution is obtained for the Weyl equation and it is argued that mass has stochastic origin. The Weyl vortex is embedded in a Gaussian wavepacket to define physical vortex. Two-vortex internal structure of electron is developed in terms of Weyl and sub-quantum Weyl vortices characterized by $\hbar$ and $f=e^2/2\pi c$ respectively. It is suggested that this model may find application in spintronics with a new perspective.
\end{abstract}
\maketitle
\section{\bf Introduction}

Spin-polarized electron beams in high energy physics have been used to probe the sub-atomic structure of matter, and elementary particles. In condensed matter, spin-polarized transport, in a way, was anticipated in 1936 by Nevill Mott \cite{1}. It has become a subject of intense research in the recent years. Immense scope for spintronics in microelectronics, material science and quantum information technology is being explored. The discovery of giant magnetoresistance (GMR) in the alternating layers of ferromagnetic metals (FM) and nonmagnetic metals (NM), and the spin-polarized tunneling (SPT) provided a great impetus to the spintronics \cite{2,3,4}. The most remarkable fact, not sufficiently highlighted in the literature, is that spintronics essentially deals with the intrinsic magnetic moment of electron; Prinz used the more realistic word magnetoelectronics for this field in 1995 \cite{2}. Spin accumulation, diffusion and transport in general signify the control and the manipulation of the spin magnetic moment of the electron $\mu_B$ in whch electric charge and electron spin are inseparably linked
\begin{equation}
\mu_B = \frac{e \hbar}{2m_e c}
\end{equation}
To put this point in perspective recall the experiments that measure the magnetic moment \cite{5}. Classically magnetic moment of a charged rotating object, whether a point particle or a rigid body, is intrinsically related to its angular momentum. In quantum theory, postulating spin angular momentum of $\hbar/2$ for the electron following Uhlenbeck-Goudsmit hypothesis or assuming Dirac equation, once electric charge is introduced the presence of spin manifests through magnetic moment (1). Therefore, an issue of fundamental importance in spintronics is the nature of charge, spin and their relationship.

In a more direct way one may ask whether electric charge and electron spin could be separated and independently manipulated. In some theoretical models of superconductivity based on one-dimensional Hubbard model it is speculated that there exist two kinds of excitations: electrically neutral fermions called spinons, and spinless charged objects called holons. Spinons and holons have different Fermi velocities making it possible to separate charge and spin transport. An early experiment gave evidence for spin and charge separation \cite{6}. If one looks at a fundamental level then the quantum description of electron has to be considered. Since Dirac equation is the most acceptable representation an important property of Dirac equation may be mentioned: it admits separated charge center and center of mass, for a critical discussion and early references see Barut and Bracken \cite{7}.
What is the relationship of this property of the Dirac equation with the separate charge and spin transport in matter? It seems this question has not been asked in the literature. 
In the conventional electronics and the classical electrodynamics electron spin and spin magnetic moment $\mu_B$ have no role; the later enters into the picture only through the magnetic properties of the matter. It seems rather strange that in spite of a vast literature on the Dirac equation the correspondence between Maxwell-Dirac and Maxwell-Lorentz equations has not been investigated except a recent work \cite{8}. In the present paper we make this correspondence more definitive and study its novel implications on spintronics. Two-vortex internal structure of the electron \cite{9} throws light on charge and spin transport in a unifying idea \cite{10} that charge also has origin in the fractional spin.

The paper is organized as follows. To elucidate the nature of the problem emanating from spintronics we first discuss aspects of spin transport in the next section. The proposed new approach to the Maxwell-Dirac equations is developed in Section 3. Physical implications on the spintronics are discussed in Section 4, and concluding remarks constitute the last section.

\section{\bf Physics of spin transport}

The physics of magnetic materials, the well-known ferromagnetic elements Fe, Ni and Co, antiferromagnetic like Cr, and many new exotic novel materials is believed to originate in the microscopic quantum theory of electron spin/magnetic moment, and their interactions. In GMR, layers of FM-NM, for example, Fe-Cr and Co-Cu are used. Hybrid structures, replacing NM by semiconductors, FM-NMS, have also been studied in semiconductor spintronics. Discovery of ferromagnetic semiconductor (FS) $Ga_{1-x}Mn_xAs$ has led to new devices: recently FS has been used as spin injector and spin detector on a high mobility 2-dimensional electron gas formed at the $Ga_{1-x}Mn_xAs-GaAs$ interface \cite{11}. The experimental results seem to disagree with the standard drift-diffusion theory motivating the authors \cite{11} to put forward ballistic spin transport hypothesis: a mean free path comparable to the device dimensions results into ballistic motion rather than the diffusive one. This hypothesis reminds us the collisionless pure ballistic transport directly observed in submicron GaAs devices in 1985 \cite{12}, and the controversy surrounding this \cite{13}. A viewpoint \cite{14} on \cite{11} brings out some intriguing questions related with it. Chen and Zhang \cite{15} develop spin and directional dependent local chemical potential approach to the spinor Boltzmann transport equation (BTE) to study ballistic spin transport. Note that the momentum and spin relaxation times in the BTE assume the existence of local quasi-equilibrium, and one could introduce equilibrium and nonequilibrium parts of the distribution function, see Eq.(3) in \cite{15}. A simpler idea could be that of postulating two ensembles of partial local quasi-equilibrium and ballistic carriers, termed lucky electrons in \cite{13}. Spin Hall effect (SHE), SPT and ballistic spin transport show the interplay between spin, magnetic moment, charge conduction and the electromagnetic fields. The role of spin-orbit coupling also has varied degree of significance in spintronics. The microscopic mechanism of various condensed matter systems in the context of spintronics is not the purpose of the present work; we limit ourselves to the macroscopic phenomenology of spin dynamics and electrodynamics.

In the classical Maxwell-Lorentz theory the Drude model of the conductivity has proved to be immensely useful
\begin{equation}
\sigma = \frac{e^2 n \tau}{m_e}
\end{equation}
The formal structure (2) is preserved in the quantum theory such that the free electron mass is modified to the effective mass in an energy band of the solid, and the relaxation time $\tau$ is calculated from quantum theory of scattering \cite{1}. For a transition metal Mott suggests the following formula
\begin{equation}
\sigma_M = \frac{e^2n_s \tau_s}{m_s} + \frac{e^2 n_d \tau_d}{m_d}
\end{equation}
Here suffixes 's' and 'd' refer to s- and d- orbit electrons respectively, and n is the charge carrier number density. The spin-dependent conduction is understood in terms of two Brillouin zones in the single-particle approximation. Band structure of a FM splits into spin up and spin down bands \cite{3}, and one treats two component current conduction corresponding to two spin directions. The spin-mixing can arise from spin-flip scattering between two current channels through the momentum exchange via electron-magnon collisions. Note, however that spin-lattice interaction and spin-orbit coupling are responsible for the relaxation of spin accumulation.

In fact, spin-orbit coupling induces spin-polarized currents in SHE, and this is also traced to the early work of Mott  \cite{16}. An unpolarized electron beam scattering with an unpolarized target results into a spatial separation of polarized electrons due to spin-orbit interaction. Dyakonov and Perel \cite{17} predicted spin accumulation on the surface of a sample using this mechanism where spin-layer thickness is determined by spin-diffusion. The basic approach is to assume a spin-density continuity equation for spin-density vector ${\bf S}$; spin-flux tensor $Q_{ij}$ and the spin relaxation time $\tau_{spin}$ are introduced to write
\begin{equation}
\frac{\partial S_i}{\partial t} + \frac{\partial Q_{ij}}{\partial x_j} + \frac{S_i}{\tau_{spin}} =0
\end{equation}
The tensor $Q_{ij}$ is assumed to comprise of three terms: spin-drift, spin-diffusion and spin-orbit interaction. In the second paper \cite{18} the effect of magnetic field is considered generalizing Eq.(4), and the phenomenological parameters are obtained in terms of the scattering amplitude. SHE has been experimentally observed in semiconductors and metals since then \cite{19}.

Chudnovsky \cite{20} formulates an analogue of Drude model for SHE. The starting point of his model is the nonrelativistic form of Dirac Hamiltonian for relativistic quantum electron. Assuming the effect of static crystal potential, impurities/imperfections potential and the external potential are embodied in the potential $U({\bf r})$ a classical approach is employed to derive Newton's equation of motion. Introducing relaxation time the Drude model is arrived at: theoretical calculations seem to compare fairly well with the experimental data on spin Hall conductivity.

The expression for the current density in terms of charge conductivity $\sigma_c$ and spin Hall conductivity $\sigma_{sh}$ derived in \cite{20} is
\begin{equation}
{\bf J} = \sigma_c {\bf E} + \sigma_{sh} ({\bf \xi} \times {\bf E})
\end{equation}
\begin{equation}
\sigma_c = \frac{e^2 n \tau}{m}
\end{equation}
\begin{equation}
\sigma_{sh} =\frac{\hbar e^3 n \tau^2}{2 m^3 c^2} A
\end{equation}
The spin polarization vector of the electron fluid ${\bf \xi}$ has the magnitude between 0 and 1 defined by
\begin{equation}
\xi = \frac{n_+ -n_-}{n}
\end{equation}
Here $n_+(n_-)$ denotes the carrier number density for spin parallel (antiparellel) to ${\bf \xi}$, and the total charge carrier density is $n=n_+ +n_-$. The constant A in (7) for a specific case is calculated to be
\begin{equation}
A = \frac{4\pi}{3} Z e n_o
\end{equation}
where -Ze and $n_0$ are the ionic charge and number density respectively. The author argues towards the end of the paper that both effective mass and free electron mass have to be taken into account: in the Bohr magneton it has to be $m_e$ while for orbital motion and the kinetic energy term in the Hamiltonian it has to be the effective mass. Ultimately, for metals $Zn_0 ~ \rightarrow n$, and only $m_e$ appears in the expression
\begin{equation}
\sigma_{sh} = \frac{2\pi \hbar}{ 3 m_e c^2} \sigma_c^2
\end{equation}
The calculation of spin Hall conductivity is carried out using the expression (10).

To summarize: Mott's discussion on expression (3) and Fert's exposition in the light of new advances, and Chudnovsky's approach underline the significance of two issues. First one concerns the importance of effective mass and relaxation time. The second is the role of quantum theory whether one has to consider Pauli-Schroedinger or Dirac equation. It may be asked if two length scales, namely, the electron charge radius $r_e=\frac{e^2}{m_e c^2}$ and the Compton wavelength $\lambda_c =\frac{\hbar}{m_e c}$ have fundamental role in the charge and spin transport. These issues bring out the necessity to understand the classical limit of the Maxwell-Dirac equation.

\section{\bf Maxwell-Dirac theory: charge and spin currents}

The Maxwell-Lorentz theory is the Maxwell field equation and a relativistic generalization of the Newton-Lorentz equation of a point electron
\begin{equation}
\partial_\mu F^{\mu\nu} =J^\nu
\end{equation}
\begin{equation}
m\frac{dv^\mu}{d\tau} =e v_\nu F^{\mu\nu}
\end{equation}
Here the velocity 4-vector $v^\mu =\frac{dx^\mu}{d\tau}$, and $\tau$ is proper time. Let us stress the fact that the formal structure of Eqs. (11) and (12) is abstracted from the experiments. Once it became known that charge, mass and spin are the observed physical attributes of the electron, there would have been a natural question whether the spin current or effect was hidden in this set of equations or these were incomplete. Logically both possibilities are allowed. One may reason that the experiments lacked the precision to detect the presence of electron spin when the Maxwell-Lorentz theory was developed, and therefore this description is incomplete. Or, the spin effect was implicitly present in the experiments hence it is hidden in the theory. Here it may be mentioned that there exists an approach in which the Maxwell equation(11) is retained as such, and an equation of motion for spin vector $S^\mu$ in analogy to Eq.(12) is discussed \cite{21}. The continuity equation for spin vector (4) in Dyakonov-Perel work may be viewed in this spirit.

Now quantum electrodynamics (QED) is the most successful modern theory in terms of the empirical precision tests. The experimental value of electron magnetic moment is not exactly $\mu_B$, and the QED calculated value in the power of $\alpha =\frac{e^2}{\hbar c}$ explains the anomalous part extremely well. Neglecting higher order terms electron magnetic moment in QED is calculated to be
\begin{equation}
\mu_e = \mu_B [1+\frac{\alpha}{2\pi}]
\end{equation}
On the other hand, single particle Dirac theory predicts spin magnetic moment given by expression (1). Dirac's own work and many investigations in the past have revealed intriguing aspects of the Dirac equation, some of them, e. g. zitterbewegung, continue to be puzzling even today. Though the theoretical origin of (13) belongs to QED, we propose a paradigm shift: treat $\mu_e$ as a physical attribute of the electron and seek alternative explanation \cite{22}. A new interpretation of $\mu_e$ is possible rearranging expression (13) in the following form
\begin{equation}
\mu_e =\frac{e}{mc} [\frac{\hbar}{2} +\frac{f}{2}]
\end{equation}
\begin{equation}
f=\frac{e^2}{2\pi c}
\end{equation}
Could we re-interpret $\frac{f}{2}$ as additional spin of electron? The standard practice is to express angular momentum in terms of Planck constant, if we follow this the spin $\frac{f}{2}$ is fractional. We suggest that the alternative interpretation, though a radical one, deserves serious consideration. For this purpose Maxwell-Dirac theory is analyzed here.

Practical utility of the nonrelativistic approximation to the Dirac equation, and treating Maxwell field semiclassically have been known since long. In the preceding section the usefulness of the nonrelativistic Dirac Hamiltonian for SHE \cite{20} has been discussed. A small, but important point is that nonrelativistic approximation does not mean classical limit, in fact, the presence of the Planck constant in Equation (1) of \cite{20} itself shows the quantum mechanical nature of the approximate Dirac Hamiltonian. However the classical approach is used in all the derivations leading to the spin Hall conductivity. The author does mention that one could instead use the Heisenberg equation of motion, but treats physical variables as c-numbers that is inconsistent with quantum theory. Of course, the well known consequences of the nonrelativistic limit of Dirac equation are the understanding of Pauli spin effect, spin-orbit coupling and Darwin term, and Dirac bispinor approximates to a 2-component spinor \cite{23}.

Rather than nonrelativistic approximation let us examine the classical limit. In the Maxwell-Dirac theory the Maxwell field equation (11) retains the formal structure in which the charge current density $J^\mu$ is replaced by the Dirac current
\begin{equation}
\partial_\nu F^{\mu\nu} =e \bar{\Psi} \gamma^\mu \Psi
\end{equation}
where bispinor $\Psi$ obeys the Dirac equation
\begin{equation}
 \gamma^\mu (i\hbar \partial_\mu - \frac{e}{c} A_\mu ) \Psi -m c \Psi =0
\end{equation}
and the electromagnetic field tensor is defined in terms of the 4-vector potential $A^\mu$
\begin{equation}
F^{\mu\nu} = \partial^\mu A^\nu - \partial^\nu A^\mu
\end{equation}
We use the standard notations, see \cite{8} for details. A remarkable property of the Dirac current $J^\mu_D =e\bar{\Psi} \gamma^\mu \Psi$ is that even in the presence of the electromagnetic interaction its form does not change; one could prove it using Dirac equation or by deriving Noether current for global and local U(1) gauge invariance of the Maxwell-Dirac action. Moreover the electric charge unit is put by hand; the actual Dirac current is probability current such that probability density is positive definite. An insightful approach to the Dirac current is the Gordon decomposition into Gordon and spin magnetization currents
\begin{equation}
 J^\mu_D = J^\mu_G +J^\mu_M
\end{equation}
\begin{equation}
 J^\mu_G = i \mu_B [\bar{\Psi} \partial^\mu \Psi - (\partial^\mu \bar{\Psi}) \Psi]
\end{equation}
\begin{equation}
 J^\mu_M =i \mu_B \partial_\nu M^{\mu\nu}
\end{equation}
\begin{equation}
 M^{\mu\nu} = \frac{1}{2} \bar{\Psi}[\gamma^\mu \gamma^\nu  - \gamma^\nu \gamma^\mu] \Psi
\end{equation}
For an interacting electron Dirac current and spin magnetization current remain unaltered in the presence of $A^\mu$ while Gordon current modifies to
\begin{equation}
 J^\mu_{G, int} = i \mu_B [\bar{\Psi} \partial^\mu \Psi - (\partial^\mu \bar{\Psi}) \Psi] -\frac{e^2}{mc^2} A^\mu \bar{\Psi} \Psi
\end{equation}
It is natural to ask which of the currents corresponds to the charge current in the Maxwell equation (11). To answer this question we proceed following the earlier study \cite{8} seeking the classical limit $\hbar \rightarrow 0$ assuming the Dirac bispinor to be
\begin{equation}
\Psi =C e^{\frac{iS}{\hbar}}
\end{equation}
and expanding 4-component column vector C in the power of Planck constant $C= C_0 +\frac{\hbar}{i} C_1 + higher~ order ~powers ~in~ \hbar$. We obtain
\begin{equation}
 J^\mu_{D, cl} = e ~\bar{C_0} \gamma^\mu C_0
\end{equation}
\begin{equation}
 J^\mu_{M, cl} =0
\end{equation}
\begin{equation}
 J^\mu_{G, cl, int} = -\frac{e}{mc} \rho (\partial^\mu S +\frac{e}{c} A^\mu)
\end{equation}
where we have defined
\begin{equation}
 \rho = \bar{C_0} C_0
\end{equation}
Note that the Gordon current in the nonrelativistic approximation corresponds to the Schroedinger current for the charge flow since the expressions for probability density and probability current density in Schroedinger theory are known to be
\begin{equation}
 \rho_s = \Psi_s^* \Psi_s
\end{equation}
\begin{equation}
 {\bf J}_s = -\frac{i\hbar}{2m} (\Psi_s^* {\bf \nabla} \Psi_s -{\bf \nabla} \Psi_s^* \Psi_s) - \frac{e}{mc} {\bf A} \Psi_s^* \Psi_s
\end{equation}
Here $\Psi_s$ is Schroedinger wave function. Therefore the structure of (27) is suggestive of it being the analogue of $J^\mu$ in the Maxwell equation (11). In the quantum theory it has to be pointed out that the probability interpretation is correct for the Dirac current, but not for Gordon current. Thus the classical limit of Dirac current $e \bar{C_0} \gamma^0 C_0$ differs from $\rho$ given by expression (28)
for the Gordon current.

Having discussed the classical correspondence the crucial step is to prove the consistency between Eqs. (11) and (16). Departing from the conventional approach we introduce modified charge current density and the electromagnetic field tensor
\begin{equation}
J^\mu_m =J^\mu_{G,int}
\end{equation}
\begin{equation}
F^{\mu\nu}_m =F^{\mu\nu} -i \mu_B M^{\mu\nu}
\end{equation}
such that
\begin{equation}
\partial_\nu F^{\mu\nu}_m =J^\mu_m
\end{equation}
An important property of the Gordon decomposition (19) is that each of the currents separately satisfies the continuity equation $\partial_\mu J^\mu_G=0$ and $ \partial_\mu J^\mu_M=0$.  This property would ensure the mathematical consistency of the proposition (31) to (33).

What is the physical significance of our proposition? First, the charge current (31) hides the electromagnetic potential and the spin magnetization tensor is hidden in the electromagnetic field tensor. Secondly the Compton wavelength of the electron in the first term, and the electron charge radius in the second term of the Gordon current (23) signify the importance of two length scales in the charge current. Alternatively, in terms of the spin both $\frac{\hbar}{2}$ and $\frac{f}{2}$ have significance in the current $J^\mu_m$. 

Third one is a subtle point. To derive the Gordon decomposition one has to use the Dirac equation with nonzero mass term. For a massless particle the Dirac equation splits into the Weyl equations for the left-handed $\Psi_L$ and right-handed $\Psi_R$ two-component Weyl spinors
\begin{equation}
i \hbar {\bf \sigma}.{\bf \nabla} \Psi_L = \frac{i\hbar}{c} \frac{\partial \Psi_L}{\partial t}
\end{equation}
\begin{equation}
i \hbar {\bf \sigma}.{\bf \nabla} \Psi_R =- \frac{i\hbar}{c} \frac{\partial \Psi_R}{\partial t}
\end{equation}
Comparing Dirac equation (17) with the set of Weyl equations (34)-(35) it is easily noticed that the Planck constant could be cancelled in the later while due to mass term in the Dirac equation it cannot be factored out. Since $f$ has the dimension of action/angular momentum, a sub-quantum description is envisaged \cite{9} such that the sub-quantum Weyl equations are given by
\begin{equation}
i f {\bf \sigma}.{\bf \nabla} \Psi_L^s = \frac{if}{c} \frac{\partial \Psi_L^s}{\partial t}
\end{equation}
\begin{equation}
i f {\bf \sigma}.{\bf \nabla} \Psi_R^s =- \frac{if}{c} \frac{\partial \Psi_R^s}{\partial t}
\end{equation}

Attention is drawn to an important aspect in the Maxwell-Lorentz theory: the charge unit $e$ can be factored out in the Maxwell equations and it appears as $e^2$ in the Lorentz force expression and the Lagrangian density when the fields are expressed in the geometrical unit ${length}^{-2} $. Introducing $\hbar$ in the redefined fields we have proposed a new interpretation that the electromagnetic field tensor represents angular momentum per unit area of the photon fluid \cite{10}. The present considerations on the Dirac current pave the path for a unifying picture: both charge current and spin current represent angular momentum flows. Electron mass is proposed to couple massless Weyl fermions $(\Psi_L, \Psi_R)$ and sub-quantum Weyl fermions $(\Psi_L^s, \Psi_R^s)$. Thus qualitatively a symbolic picture of the electron is as follows
\begin{equation}
e^- = \Psi^s_L ; \Psi_L,\Psi_R
\end{equation} 
\begin{equation}
e^+ = \Psi^s_R ; \Psi_L,\Psi_R
\end{equation} 
The sign of the electric charge is determined by the handedness of sub-quantum Weyl spinor, and the length scales $r_e$ and $\lambda_c$ correspond to sub-quantum and quantum states of the electron structure respectively.

\section{\bf Electron vortex and spintronics}

The microscopic theory of spin transport and technical complexities involved in spintronics are specialized topics receiving a great deal of attention. However the applicability of Dirac equation in this field has a general validity at a basic level. Interpretation of Dirac equation has been approached from diverse angles in the literature \cite{7, 24,25,26,27,28,29,30}. Physical interpretation of Dirac equation assuming a point electron leads to a counter-intuitive picture of zitterbewegung which is not observable, a length scale of the order of $\lambda_c$, and the presence of the center of charge and center of mass separated by $\lambda_c /2$. An alternative, though investigated by very few physicists, seeks an internal structure of the electron incorporating zitterbewegung and the Compton wavelength. Such efforts are reminiscent of the past electron models beginning with that of Thomson inspired by classical electron charge radius. The failure of such models does not necessarily imply the validity of a point electron, in fact, the foundational problems originating from the infinities show the limitations of the point field theories e. g. QED. It is also true that to make progress in the internal structure models a radically new idea is imperative. We have proposed two-vortex model towards this aim \cite{9} and given a qualitative picture in \cite{22}. The new analysis on the Maxwell-Dirac theory shows that the spin transport and the charge transport are physically same at the basic level both being the angular momentum flows. The role of mass, coupling massless Weyl and sub-quantum Weyl spinors, and the vortex model of the electron are proposed to have a profound implication on the spintronics. 

Though nonrelativistic limit of Dirac equation and semiclassical equations of motion are used in Chudnovsky's phenomenological model of SHE \cite{20} an interesting insight could be obtained from this work. Charge and spin conductivity depend on the mass of the electron but the significance of mass parameter to be used whether free electron mass $m_e$ or the effective mass in the material $m_{eff}$ depends on physical arguments. Chudnovsky suggests that $m^2$ in the spin-orbit term should be the product $m_e m_{eff}$, and shows that finally it is only $m_e$ that appears in the expression (10). Now if we take the limit $m_e \rightarrow 0$ in Eq.(10) then charge conductivity vanishes but $\sigma_{sh}$ could be nonzero. It may be argued that this implies pure spin transport. A recent study based on the standard theory of spin-orbit coupling analyzes the conversion between spin and charge currents by Edelstein and inverse Edelstein effects \cite{31}. Note that Dirac cone and spinor Boltzmann equation are used in this study. 

The nature of mass becomes important in this connection. Usually the classical correspondence of quantum theory is approached taking the limit $\hbar ~ \rightarrow ~ 0$. In a different approach \cite{32} the question is raised as to the consequence of taking the limit $m ~\rightarrow ~0$ in the Schroedinger equation. It is argued that this opens up the possibility of a field interpretation. Free particle Schroedinger equation for this purpose is rewritten in the form
\begin{equation}
\nabla ^2 \Psi_s = \frac{2m}{i\hbar} \frac{\partial \Psi_s}{\partial t}
\end{equation}
that reduces to the Laplace equation as $m~\rightarrow ~0$. Simplest relativistic generalization  is then
\begin{equation}
\nabla ^2 \Psi_s=\frac{1}{c^2} \frac{\partial ^2 \Psi_s}{\partial t^2}
\end{equation}
Interpreting mass as the influence of the surrounding medium a modified Schroedinger equation is obtained \cite{32}. An alternative derivation of Dirac equation using this idea was also given. What is the physical meaning of the surrounding medium? Is it akin to de Broglie's hidden thermostat \cite{33}? In an interesting paper \cite{34} Dirac equation in Weyl representation is derived assuming stochastic process for massless Weyl spinors undergoing random spin flips such that the mass determines the rate of the flips. Once again the source of stochasticity is undefined. However the stochastic approach to quantum theory developed over past many decades \cite{35} indicates that some of the arguments have a potential to offer an alternative to quantum mysteries.

We propose to develop the electron model represented by the symbolic equivalence (39) introducing two new ideas: deriving a vortex solution to the Weyl equation, and embedding the vortex in a host Gaussian wavepacket of stochastic origin. It is found that there exists a singularity in the Weyl equation. The nature of singularity is that of a vortex; in fact, it is a phase singularity given a fluid dynamical analogy \cite{10}. Phase singularity is defined by a vanishing field on the axis where phase is indeterminate.
Vortex represents a phase singularity, and in the fluid dynamical interpretation phase is a velocity potential. A line singularity around which the flow takes place in concentric circles is a vortex line.

In the standard treatment, neutrino Weyl equation is solved assuming infinite plane wave
\begin{equation}
\Psi = e^{-\frac{i}{\hbar} (E t -{\bf p}.{\bf x})} u(p)
\end{equation}
Nonvanishing solutions exist if $E=\pm |{\bf p}| c$. The sign of energy corresponds to a definite helicity. Physical electron neutrino has spin antiparallel to the direction of propagation and satisfies Eq.(34) for positive energy solution. We seek a vortex solution to Weyl equation, let us consider Eq.(35). Assuming the spinor to be $\left(\begin{array}{cc} \Phi\\ 0 \\ \end{array} \right)$ this equation becomes
\begin{equation}
i \hbar \left(\begin{array}{cc} \frac{\partial\Phi}{\partial z} +\frac{1}{c} \frac{\partial \Phi}{\partial t}\\ \frac{\partial\Phi}{\partial x} +i \frac{\partial \Phi}{\partial y} \\ \end{array} \right)=0
\end{equation}
A general solution of Eq.(43) in cylindrical coordinates $(r, \theta, z)$ is obtained to be
\begin{equation}
\Phi =\Phi_0 r^l e^{i l \theta} e^{-\frac{i}{\hbar}(E t -p_z z)}
\end{equation}
where $\Phi_0$ is a constant amplitude and the azimuthal index $l$ determines the order of the singularity: $\Phi$ vanishes at $r=0$ and phase $l\theta$ becomes undefined. This topological defect is a phase vortex. However just like infinite plane wave that extends over whole space the solution (44) is unphysical as the field grows in the transverse plane to infinity as $r \rightarrow \infty$; in fact, this behavior is worse than the plane wave.

To solve this problem two physical arguments are used. First we seek guidance from the physics of the optical vortices \cite{10}. Experimentally it is known that the electric field in laser beams could possess a phase singularity in a host wavepacket of Hermite-Gaussian (HG) and Laguerre-Gaussian (LG) modes \cite{36}. Curiously HG and LG modes are the solutions of the paraxial wave equation, and these are inconsistent with the Maxwell field equations \cite{10,37}. Analogous approach for the Weyl equation could be to use the paraxial wave equation for the Weyl spinor since it also satisfies the d'Alembert wave equation. However we adopt a different approach: the exact solution (44) is embedded in a Gaussian wavepacket that has origin in the external space. What is this 'external space' ? Let us recall that de Broglie in the thermodynamic argument \cite{33} speculates that, 'a particle, even when isolated from a complete macroscopic body, is constantly in thermal contact with a kind of thermostat residing in what we shall call the void'. The quantum vacuum of the modern quantum field theories is known to have observable physical effects, therefore we may identify this with the external space. Note that for the zero-point field the ground state wavepacket of a harmonic oscillator is a Gaussian function; the Gaussian is a minimum uncertainty wavepacket. The importance of zero-point radiation field in the stochastic electrodynamics is also recognized \cite{35,38}. Secondly the normal or Gaussian distribution is one of the most useful continuous probability distributions. Hence it is reasonable to assume Gaussian function for a physical vortex: the vortex field (44) is implanted in a Gaussian profile in the transverse plane to obtain
\begin{equation}
\Phi_{phys} = e^{-\frac{r^2}{R^2}} ~ \Phi
\end{equation}
We have not included the normalization factor in Eq.(45).

Phase vortex solutions of the kind (45) would immediately follow for the Weyl and sub-quantum Weyl equations (34)-(37). Denoting the Weyl and sub-quantum Weyl vortices by C and O respectively the dimensionless vortex strengths for them are given by
\begin{equation}
\Gamma_g =\frac{1}{2}
\end{equation}
\begin{equation}
\Gamma_e =\frac{\alpha}{4\pi}
\end{equation}
The sign of the charge, i. e. negative for electron, corresponds to the anti-vortex $O^-$ with strength $\Gamma_e$  and the spin vortex $C$ has strength $\Gamma_g$. The conventional picture of spin up and spin down point electron is altered to inequivalent internal structures $(O^-,C^+)$ and $(O^-,C^-)$ respectively. The internal constituents of the electron travel with the velocity of light independently in the longitudinal direction (along z-axis), and have azimuthal flow in the transverse plane that determines the angular momentum: here the physical picture is similar to Huang's \cite{24} in which intrinsic spin is attributed to the orbital angular momentum of electron in a circular orbit of radius $\lambda_c$. Since the vortices are embedded in a Gaussian wavepacket the stochastic process results into an interaction between them. The half-width parameter of the wavepacket and the separation between the vortices determines the interaction. The study on the vortex-vortex interaction in optical fields \cite{39} throws light on a plausible mechanism for the electron vortices. Let the vortex C be positioned on the z-axis, and the vortex O is at ${\bf r}_0 =(x_0,y_0)$ at a radial distance $\lambda_c$. The vortex O would undergo a spiral motion such that its momentum along z-direction $p_z$ acquires a tilt at some angle, say $\theta_0$ from the z-axis. Transverse component $p_z sin \theta_0$, assuming $p_z =m_e c$ and $\theta_0 =\alpha/2\pi$ gives the orbital angular momentum 
\begin{equation}
p_z \frac{\alpha}{2\pi} \lambda_c =\frac{e^2}{2\pi c} =f
\end{equation}
where $sin \theta_0$ approximates to $\theta_0$ for small tilt. This naive argument shows that the fractional spin $f$ is linked to the orbiting vortex $O^-$. In contrast to a counter-intuitive consequence of a point electron in Dirac equation having two centers \cite{7,25} the present model gives a concrete physical explanation of both Compton wavelength and electron charge radius.

Spintronics and topological matter seem to be important for testing the vortex electron model: the main idea is that the topology of the internal structure of electron manifests in the topological properties of the matter, and spin transport and charge transport have fundamentally same physical nature, i. e. angular momentum flow. For a quantitative prediction Eqs (31) -(33) have to be related with the spin current phenomenology \cite{11,17,18,19,20,31}. To understand topological implications a rigourous treatment of vortex dynamics is required. However it could be anticipated that the host Gaussian wavepacket in matter will have different value of R and it will have temperature dependence since the lattice vibrations become important. As a consequence the interaction between the vortices O and C will be changed, and in certain situations charge vortex O and spin vortex C could travel with velocity of light independent of each other. Thus the charge and spin separation implies the motion at light speed in this model, and it should be testable.

\section{\bf Conclusion}

The present work brings out salient features of the Maxwell-Dirac theory by taking the classical limit. Its application in spintronics is pointed out. A vortex solution is obtained for the Weyl equation. Two-vortex model of electron \cite{9} is developed with new ingredients of the Weyl and sub-quantum Weyl vortices.

There are at least two directions for further development of these ideas. The theory of vortex dynamics and vortex-vortex interaction in the present model requires a rigorous treatment. The second issue is regarding the incorporation of mass term and establishing its equivalence with the Dirac equation. Though we anticipate the usefulness of the previous works \cite{32,34} in this objective we have not been able to do it at present. A simple method generalizing, for example, (35) and (36) as coupled equations using $m_e$ does not work since Eq.(36) becomes
\begin{equation}
i \hbar {\bf \sigma}.{\bf \nabla} \Psi_L^s = \frac{i\hbar}{c} \frac{\partial \Psi_L^s}{\partial t}+\frac{2 \pi}{\alpha} m_e c \Psi_R
\end{equation}
It is also not clear what the stochastic process could be.

\end{document}